\documentclass[twocolumn,superscriptaddress,10pt]{revtex4-1}
\usepackage[utf8]{inputenc}
\usepackage{amsmath}
\usepackage{amssymb}
\usepackage{graphicx}
\usepackage{subfigure}
\usepackage{enumerate}
\usepackage{textcomp}
\usepackage{chemfig}
\newcommand{\QE}{\mbox{Quantum-Espresso}}
\newcommand{\Elk}{\mbox{Elk}}
\newcommand{\cminv}{$\textrm{cm}^{-1}$}
\newcommand{\mkm}{\textmu{}m}
\newcommand{\mks}{\textmu{}s}
\newcommand{\Biz}{$\textrm{Bi}^{0}$}
\newcommand{\Bipi}{$\textrm{Bi}^{+}$}
\newcommand{\Bipii}{$\textrm{Bi}^{2+}$}
\newcommand{\Bipiii}{$\textrm{Bi}^{3+}$}
\newcommand{\GeGe}{\mbox{$\equiv\!\textrm{Ge}\!\relbar\!\textrm{Ge}\!\equiv$}}
\newcommand{\Term}[4]{\mbox{${}^{#1}{\textrm{#2}}_{#3}{#4}$}}

\begin{document}
\title{%
Infrared luminescence \\
in Bi-doped \mbox{Ge--S} and \mbox{As--Ge--S} chalcogenide glasses and fibers
}
\author{V.~G.~Plotnichenko}
%\email{Corresponding author: victor@fo.gpi.ac.ru}
\affiliation{Fiber ics Research Center of the Russian Academy of
Sciences \\ 38 Vavilov Street, Moscow, 119333, Russia}
\affiliation{Moscow Institute of Physics and Technology \\
9 Institutskii per., Dolgoprudny, Moscow Region, 141700, Russia}
\author{D.~V.~Philippovskiy}
\affiliation{Fiber ics Research Center of the Russian Academy of
Sciences \\ 38 Vavilov Street, Moscow, 119333, Russia}
\author{V.~O.~Sokolov}
\affiliation{Fiber ics Research Center of the Russian Academy of
Sciences \\ 38 Vavilov Street, Moscow, 119333, Russia}
\author{M.~V.~Sukhanov}
\affiliation{Institute of Chemistry of High-Purity Substances of the Russian
Academy of Sciences \\ 49~Tropinin Street, Nizhny Novgorod, 603600 Russia}
\author{A.~P.~Velmuzhov}
\affiliation{Institute of Chemistry of High-Purity Substances of the Russian
Academy of Sciences \\ 49~Tropinin Street, Nizhny Novgorod, 603600 Russia}
\author{M.~F.~Churbanov}
\affiliation{Institute of Chemistry of High-Purity Substances of the Russian
Academy of Sciences \\ 49~Tropinin Street, Nizhny Novgorod, 603600 Russia}
\author{E.~M.~Dianov}
\affiliation{Fiber ics Research Center of the Russian Academy of
Sciences \\ 38 Vavilov Street, Moscow, 119333, Russia}
\begin{abstract}
Experimental and theoretical studies of spectral properties of chalcogenide
\mbox{Ge--S} and \mbox{As--Ge--S} glasses and fibers are performed. A broad
infrared (IR) luminescence band which covers the 1.2--2.3~\mkm{} range with a
lifetime about 6~\mks{} is discovered. Similar luminescence is also present in
optical fibers drawn from these glasses. Arsenic addition to \mbox{Ge--S} glass
significantly enhances both its resistance to crystallization and the intensity
of the luminescence. Computer modeling of Bi-related centers shows that
interstitial \Bipi{} ions adjacent to negatively charged S vacancies are most
likely responsible for the IR luminescence.
\end{abstract}

\maketitle

\section{Introduction}
During the last decade Bi-doped bulk glasses and optical fibers have attracted a
lot of interest due to their characteristic broadband IR luminescence in the
1--1.7~\mkm{} range which has a potential for new broadband fiber amplifiers and
lasers \cite{Dianov13}. Despite a successful demonstration of laser generation
and optical gain in the 1.15--1.55~\mkm{} range using silica-based optical
fibers with different dopants \cite{Dianov12}, the origin of active Bi-related
centers in optical materials is still controversial. IR luminescence from such
centers has been observed in different types of glasses and crystals, and the
spectral properties of infrared (IR) luminescence have similar features. The
available experimental data suggest that Bi luminescent centers have similar
origin in different hosts. Besides there are reasons to believe that there are
several active centers in the glass sample and the ratio between them depends on
the host and on production technology.

Certain properties of chalcogenide glasses \cite{Zakery03} make them, on the one
hand, promising materials for various optical devices and applications including
fiber optics, and on the other, convenient hosts for optical transitions
research, in particular for the luminescence studies. These properties include a
wide transparency range (from visible to middle-IR) that depends on glass
composition, high refractive index, and high nonlinear susceptibility. The
important features of these glasses are an ability to be doped with rare-earth
elements and promising luminescence characteristics of rare-earth active centers
\cite{Shaw01, Churbanov03, Tang08}. Low phonon energy allows radiative
transitions in the middle-IR region (with wavelengths beyond 2~\mkm). A
significant number of chalcogenide glasses shows good chemical resistance,
especially to atmospheric water. It is also possible to obtain glasses with a
wide variety of properties depending on their composition.

Chalcogenide glasses allow one to research the origin of Bi-related IR
luminescence, previously studied in the following glass systems:
GeS$_2$--Ga$_2$S$_3$--KBr \cite{Yang07}, GeS$_2$--Ga$_2$S$_3$ \cite{Dong08},
and Ga$_2$S$_3$--La$_2$S$_3$--La$_2$O$_3$ \cite{Hughes09}. In \cite{Yang07,
Dong08} bismuth in low-valence states, such as \Bipi, was supposed to be
responsible for the IR luminescence. Based on calculation results
\cite{Sokolov09}, bismuth dimers were considered in \cite{Hughes09} as possible
sources of the IR luminescence.

For our research we chose two glass systems: \mbox{Ge--S} and \mbox{As--Ge--S}.
The first was used for studying the influence of Ge/S ratio on luminescence
intensity. The second system is far more resistant to crystallization, and so it
is suitable for fiber drawing. We also compared our experimental results with
\mbox{Ga--Ge--S} glass studied earlier.
\begin{figure*}
\subfigure[]{%
\includegraphics[width=8.75cm, bb=70 290 550 770]{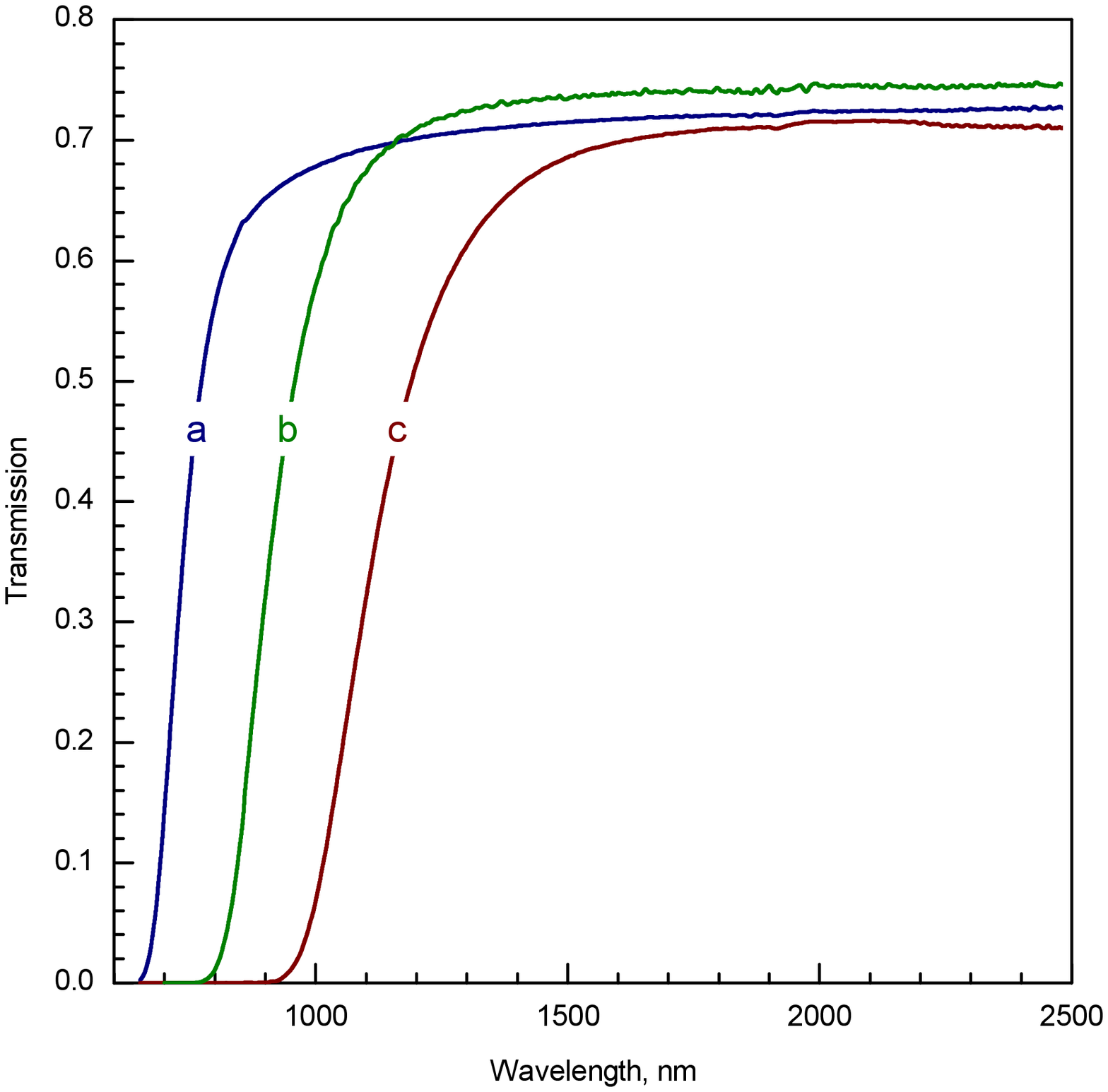}
\label{fig:1A}
}
\subfigure[]{%
\includegraphics[width=8.75cm, bb=70 290 550 770]{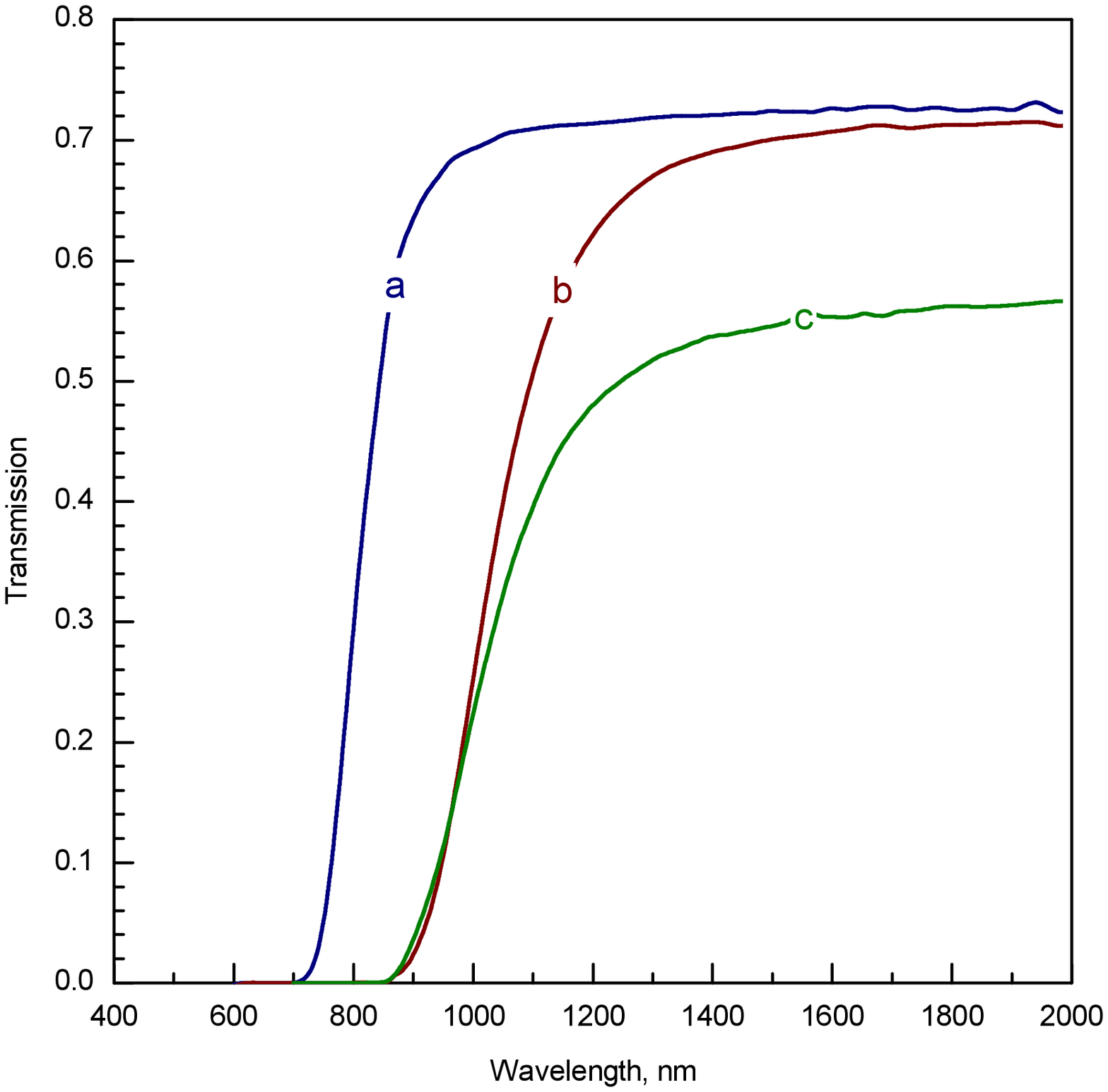}
\label{fig:1B}
} 
\subfigure[]{%
\includegraphics[width=8.75cm, bb=70 290 550 770]{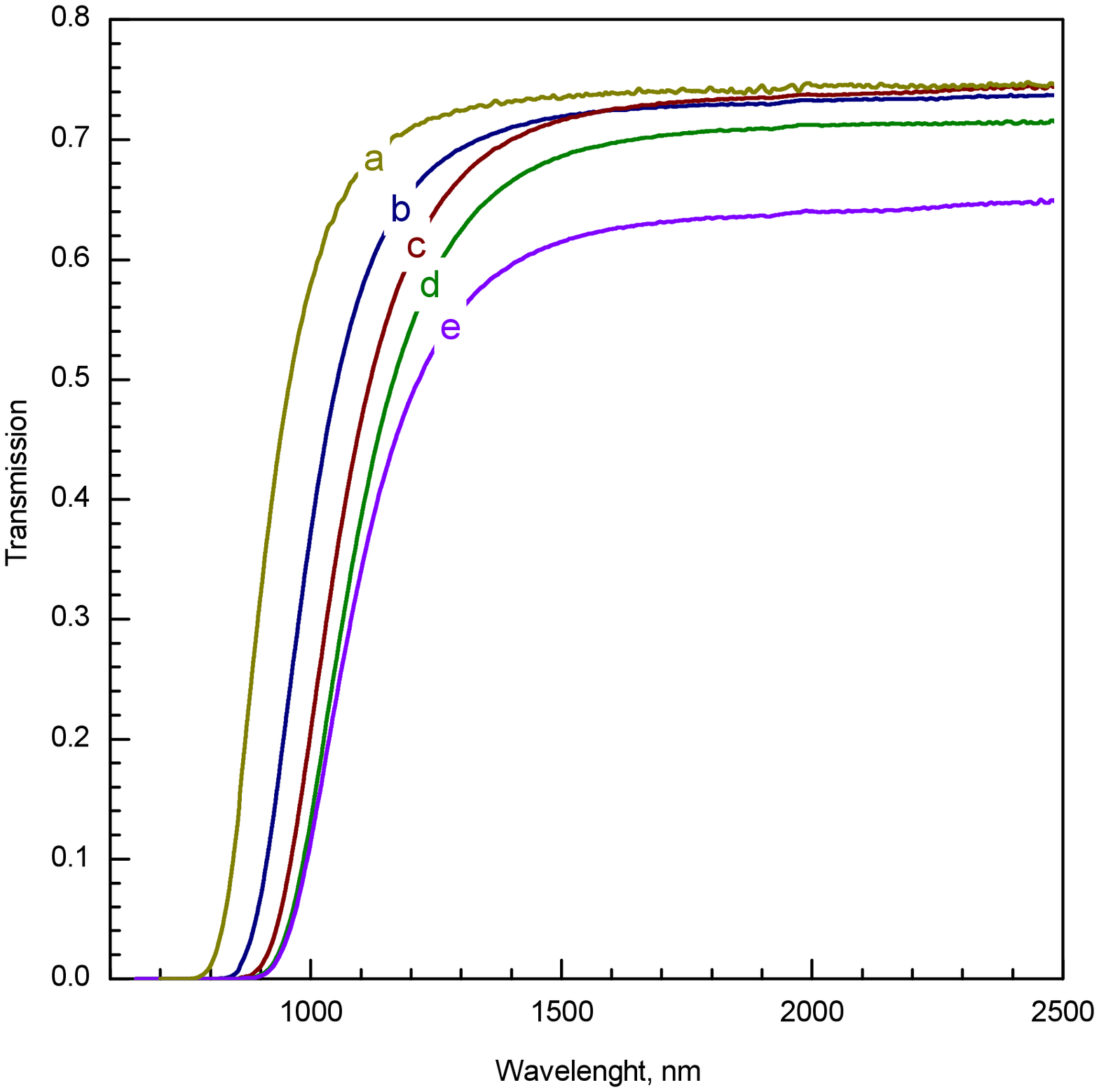}
\label{fig:1C}
}
\subfigure[]{%
\includegraphics[width=8.75cm, bb=70 290 550 770]{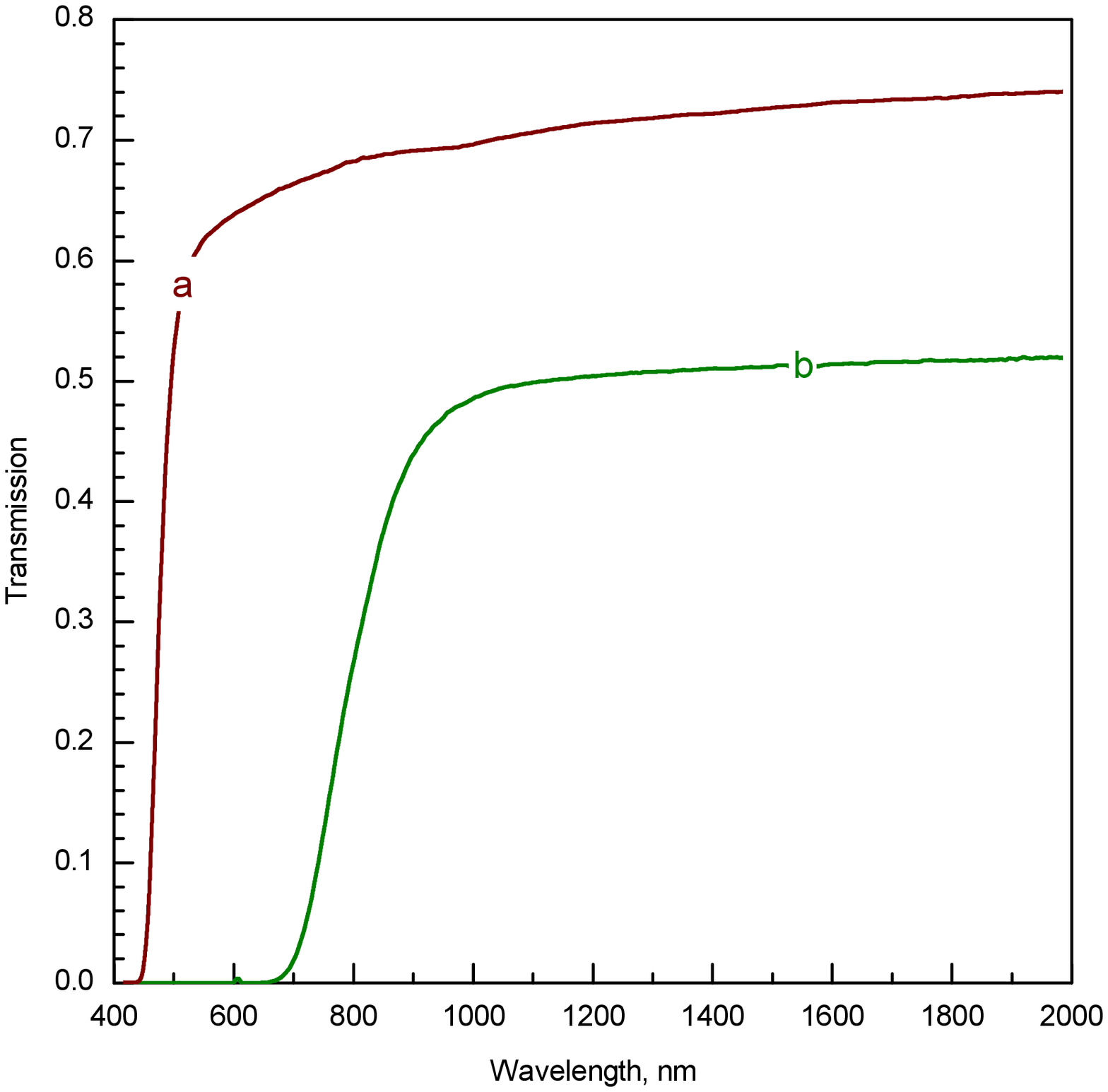}
\label{fig:1D}
}
\caption{%
Transmission spectra.
\subref{fig:1A}.~\mbox{As--Ge--S} glasses with different Bi concentration:
(a)~\mbox{As$_{10}$Ge$_{36}$S$_{54} + 0.05$\%~Bi},
(b)~\mbox{As$_{10}$Ge$_{35}$S$_{55} + 0.5$\%~Bi},
(c)~\mbox{As$_{10}$Ge$_{36}$S$_{54} + 1$\%~Bi}; \quad
\subref{fig:1B}.~\mbox{Ge--S} glasses:
(a)~GeS$_{1.35}$,
(b)~\mbox{GeS$_{1.35} + 1$\%~Bi},
(c)~\mbox{GeS$_{1.5} + 1$\%~Bi}; \quad
\subref{fig:1C}.~\mbox{As--Ge--S} glasses with different As content:
(a)~\mbox{As$_{10}$Ge$_{35}$S$_{55} + 0.5$\%~Bi},
(b)~\mbox{As$_{5}$Ge$_{38}$S$_{57} + 0.5$\%~Bi,}
(c)~\mbox{As$_{2}$Ge$_{39.2}$S$_{58.8} + 0.5$\%~Bi,}
(d)~\mbox{AsGe$_{39.6}$S$_{59.4} + 0.5$\%~Bi},
(e)~\mbox{As$_{0.5}$Ge$_{39.8}$S$_{59.7} + 0.5$\%~Bi}; \quad
\subref{fig:1D}.~\mbox{Ga--Ge--S} glasses:
(a)~\mbox{80GeS$_{2}$--20Ga$_2$S$_{3}$},
(b)~\mbox{80GeS$_{2}$--20Ga$_2$S$_{3} + 1$\%~Bi}.
}
\label{fig:1}
\end{figure*}

\section{Experiment}
Metallic Bi doped bulk glass samples (concentrations 0.05, 0.5, 1~at.\%), as
well as samples without Bi were synthesized from high-purity Ge, Ga, As, S in
evacuated quartz ampules with inner diameter 10--12~mm in a rocking muffle
furnace at 800--850~$^\circ$C. For observation of luminescence dependence on
synthesis temperature \mbox{Ge--S} samples were synthesized at 700~$^\circ$C.
The times of cooling and annealing of \mbox{Ge--S} and \mbox{As--Ge--S} glasses
were 2 and 15~hours, respectively \cite{Scripachev87}.

For spectroscopic studies we used 2--3~mm thick optically polished bulk samples
and 300~\mkm{} fibers drawn by the crucible method. The transmission spectra of
bulk samples were measured on a Perkin-Elmer Lambda~900 spectrometer.
Semiconductor laser sources with fiber output at 975~nm and 1064~nm wavelengths
were used as excitation sources. The output end of the fiber was fixed at a
focal point of the lens, and collimated beam was used for luminescence
excitation in the bulk samples. A setup consisting of a Hamamatsu P7163 InAs
photovoltaic detector, an MDR-2 monochromator, a collimator, and an SR830
Stanford Research lock-in amplifier was used to measure the luminescence
spectra. This equipment was able to obtain spectra in the 1--2.4~\mkm{} range
with a resolution of up to 4~nm. Samples were measured at 293--298~K
temperature.
\begin{figure*}
\subfigure[]{%
\includegraphics[width=8.75cm, bb=70 290 550 770]{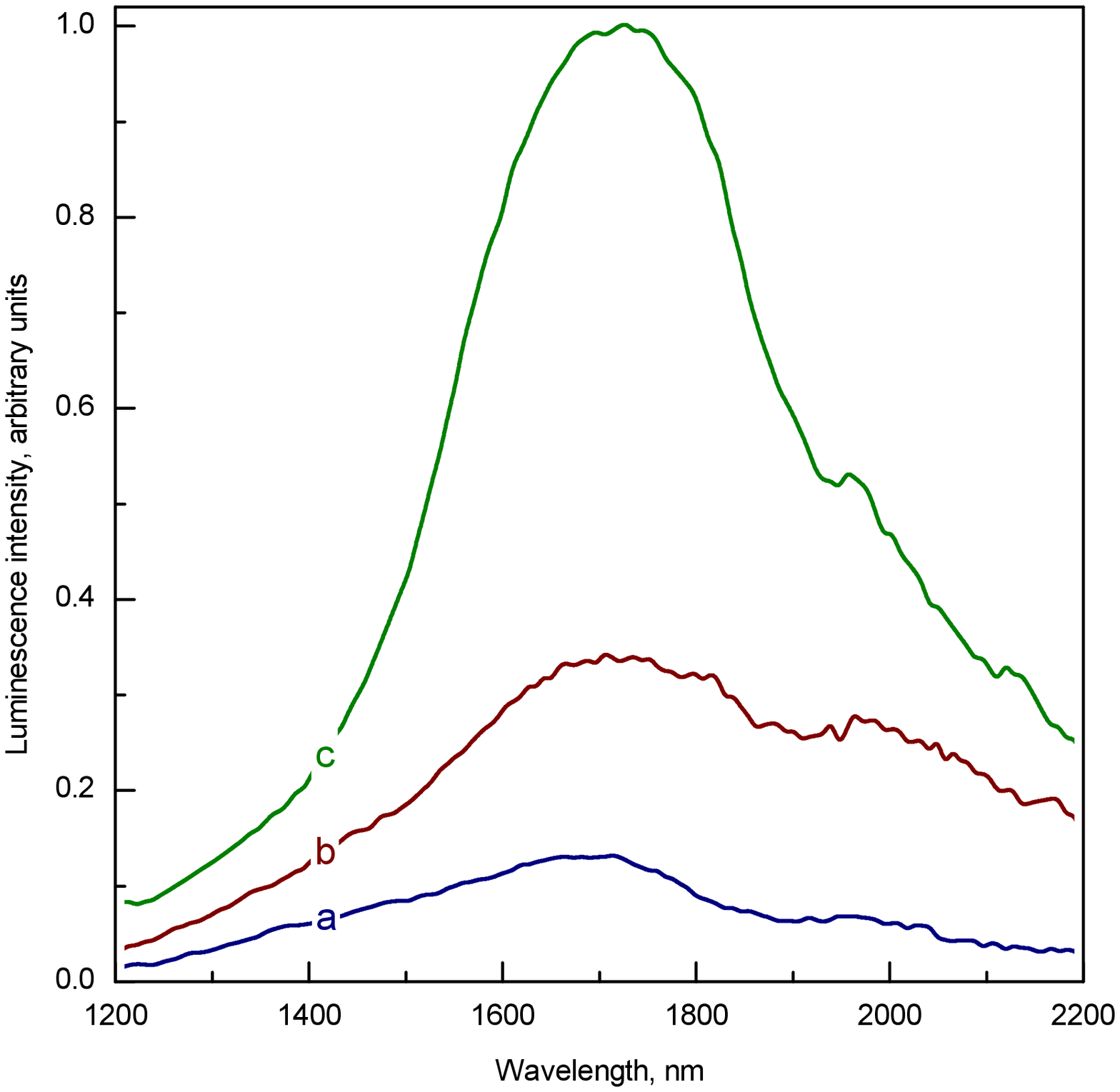}
\label{fig:2A}
}
\subfigure[]{%
\includegraphics[width=8.75cm, bb=70 290 550 770]{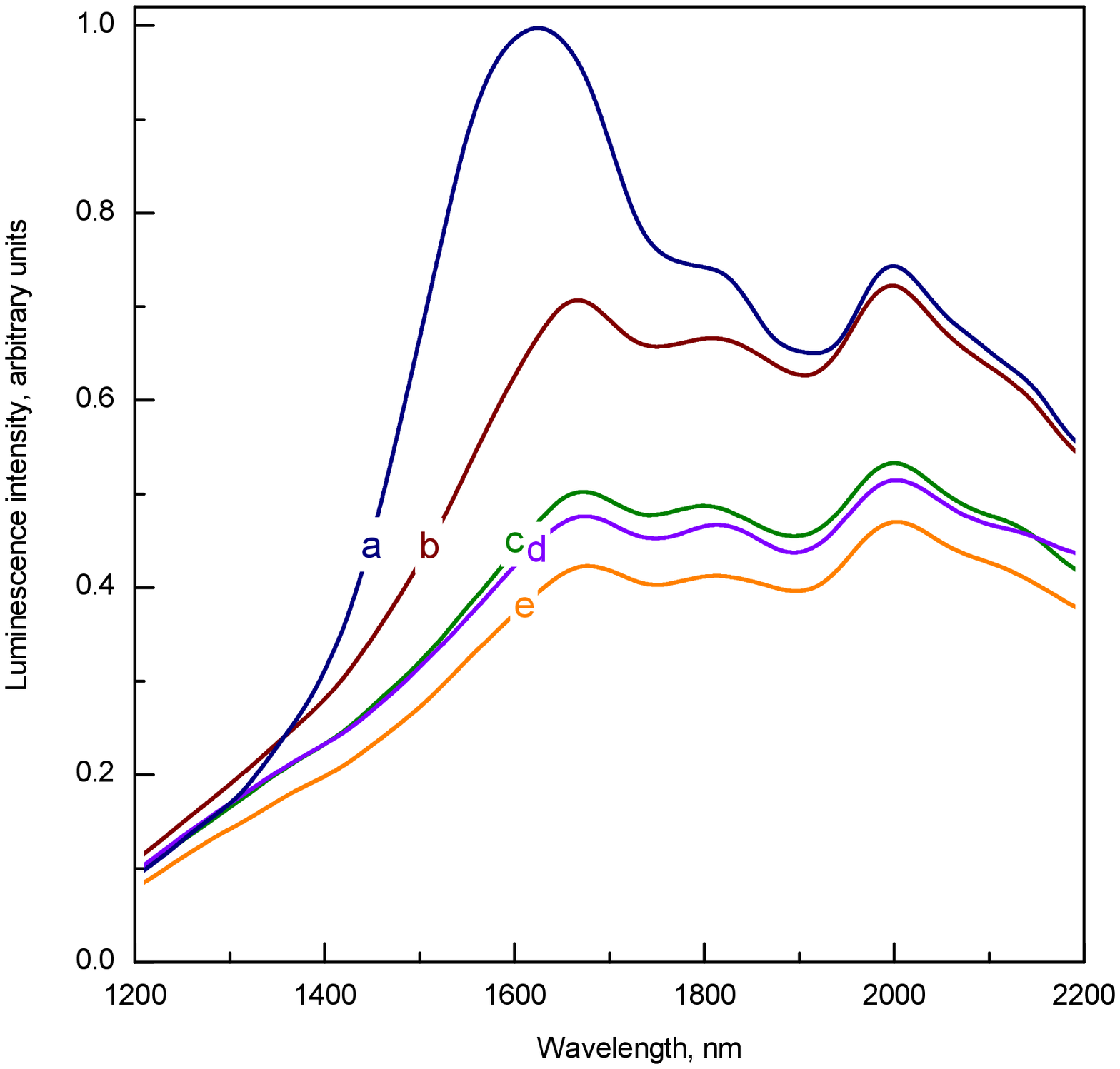}
\label{fig:2B}
}
\subfigure[]{%
\includegraphics[width=8.75cm, bb=70 290 550 770]{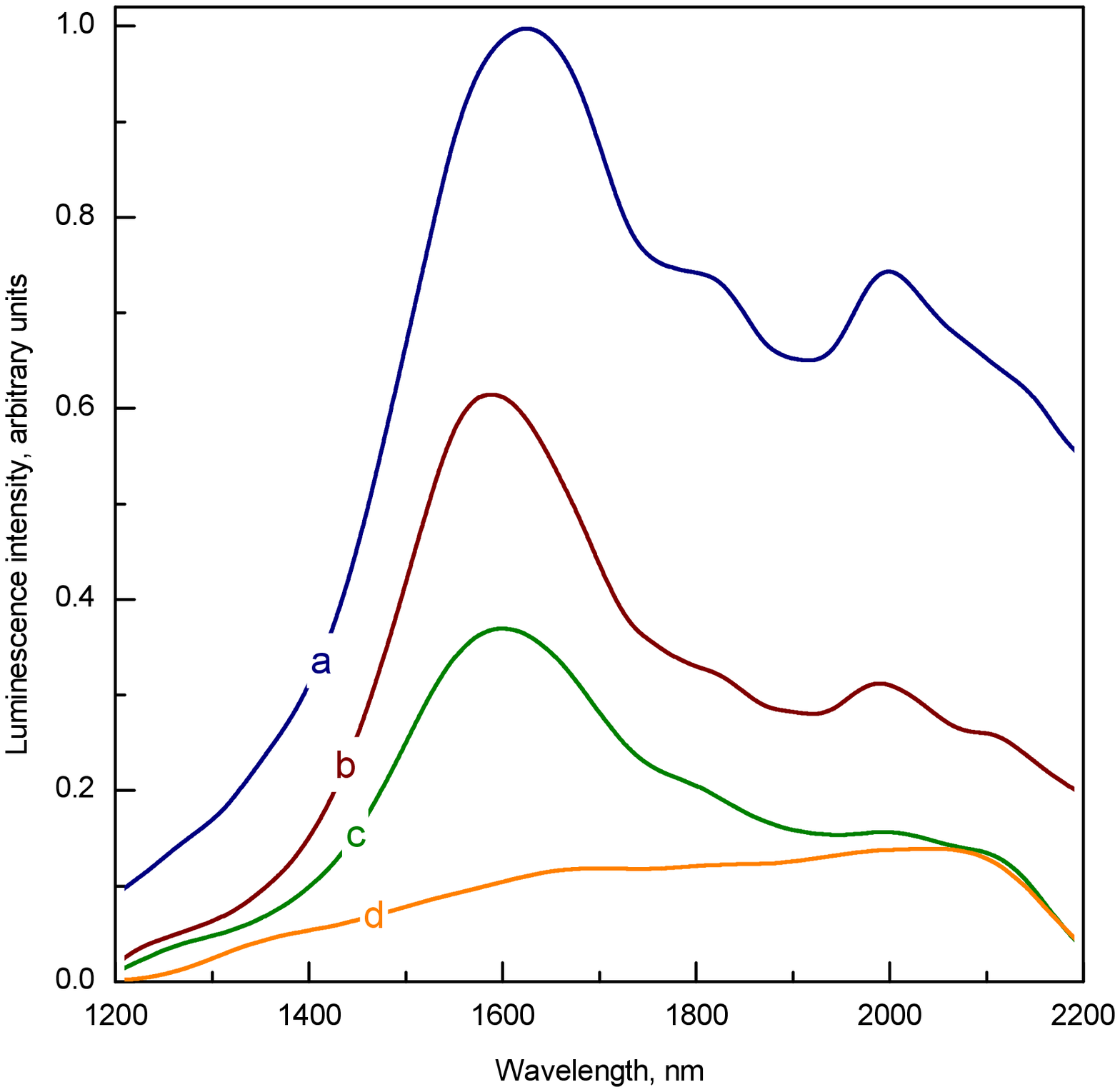}
\label{fig:2C}
}
\subfigure[]{%
\includegraphics[width=8.75cm, bb=70 290 550 770]{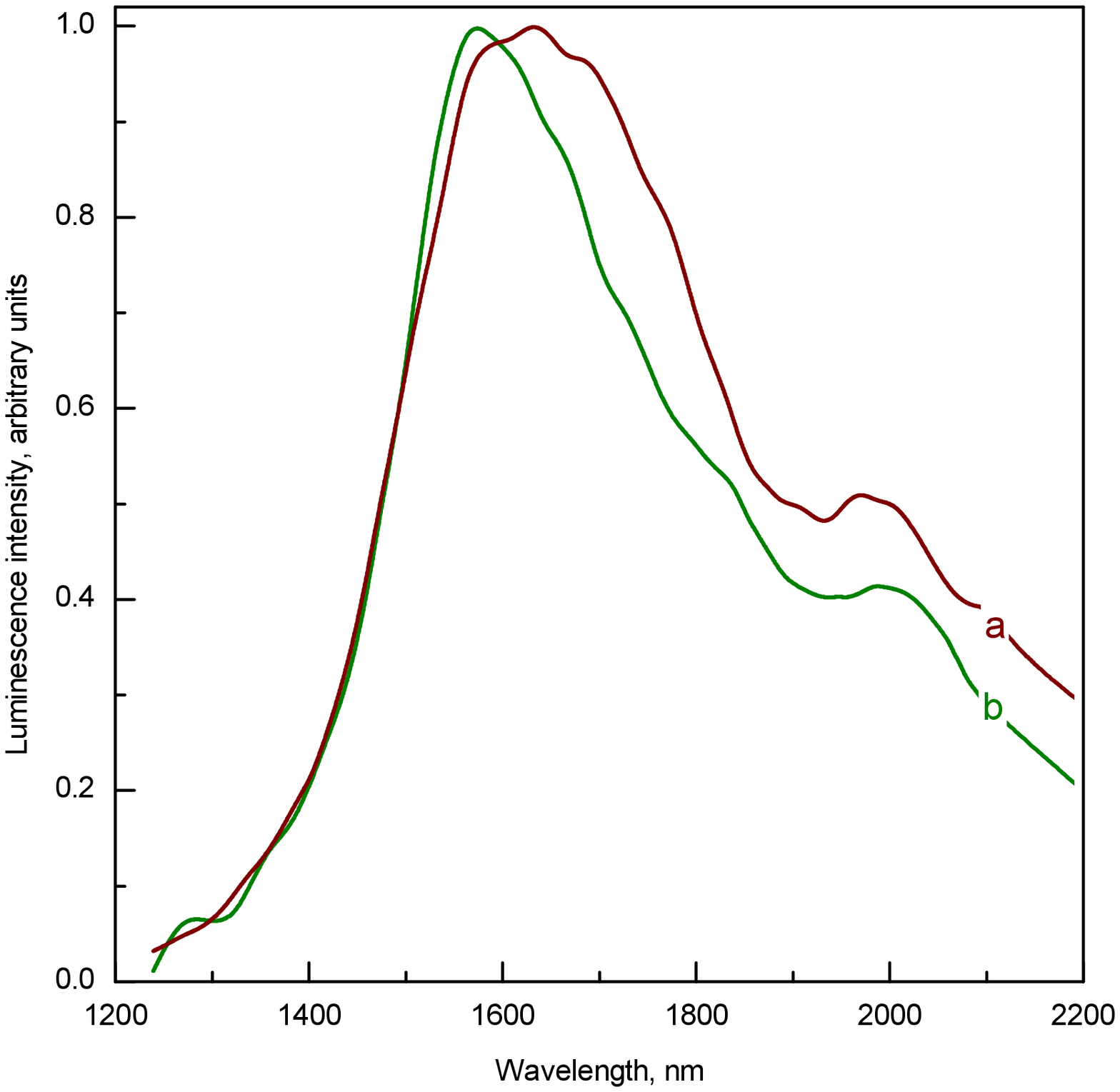}
\label{fig:2D}
}
\caption{%
Luminescence spectra.
\subref{fig:2A}.~\mbox{Ge--S} glasses (excitation at 1064~nm):
(a)~\mbox{GeS$_{1.35} + 1$\%~Bi},
(b)~\mbox{GeS$_{1.5} + 1$\%~Bi},
(c)~\mbox{As$_{10}$Ge$_{35}$S$_{55} + 0.5$\%~Bi}; \quad
\subref{fig:2B}.~\mbox{As--Ge--S} glasses with various As content (excitation at
975~nm):
(a)~\mbox{As$_{10}$Ge$_{35}$S$_{55} + 0.5$\%~Bi},
(b)~\mbox{As$_{5}$Ge$_{38}$S$_{57} + 0.5$\%~Bi},
(c)~\mbox{As$_{2}$Ge$_{39.2}$S$_{58.8} + 0.5$\%~Bi},
(d)~\mbox{AsGe$_{39.6}$S$_{59.4} + 0.5$\%~Bi},
(e)~\mbox{As$_{0.5}$Ge$_{39.8}$S$_{59.7} + 0.5$\%~Bi}; \quad
\subref{fig:2C}.~\mbox{As--Ge--S} glasses (excitation at 975~nm):
(a)~\mbox{As$_{10}$Ge$_{35}$S$_{55} + 0.5$\%~Bi},
(b)~\mbox{As$_{20}$Ge$_{30}$S$_{50} + 0.5$\%~Bi},
(c)~\mbox{As$_{10}$Ge$_{36}$S$_{54} + 0.05$\%~Bi},
(d)~\mbox{As$_{10}$Ge$_{36}$S$_{54} + 1$\%~Bi}; \quad
\subref{fig:2D}.~\mbox{As--Ge--S} fiber samples (excitation at 975~nm, the
spectra are normalized to 1):
(a)~\mbox{As$_{10}$Ge$_{35}$S$_{55} + 0.5$\%~Bi},
(b)~\mbox{As$_{20}$Ge$_{30}$S$_{50} + 0.5$\%~Bi}.
}
\label{fig:2}
\end{figure*}
\begin{figure*}
\subfigure[]{%
\includegraphics[width=8.75cm, bb=70 290 550 770]{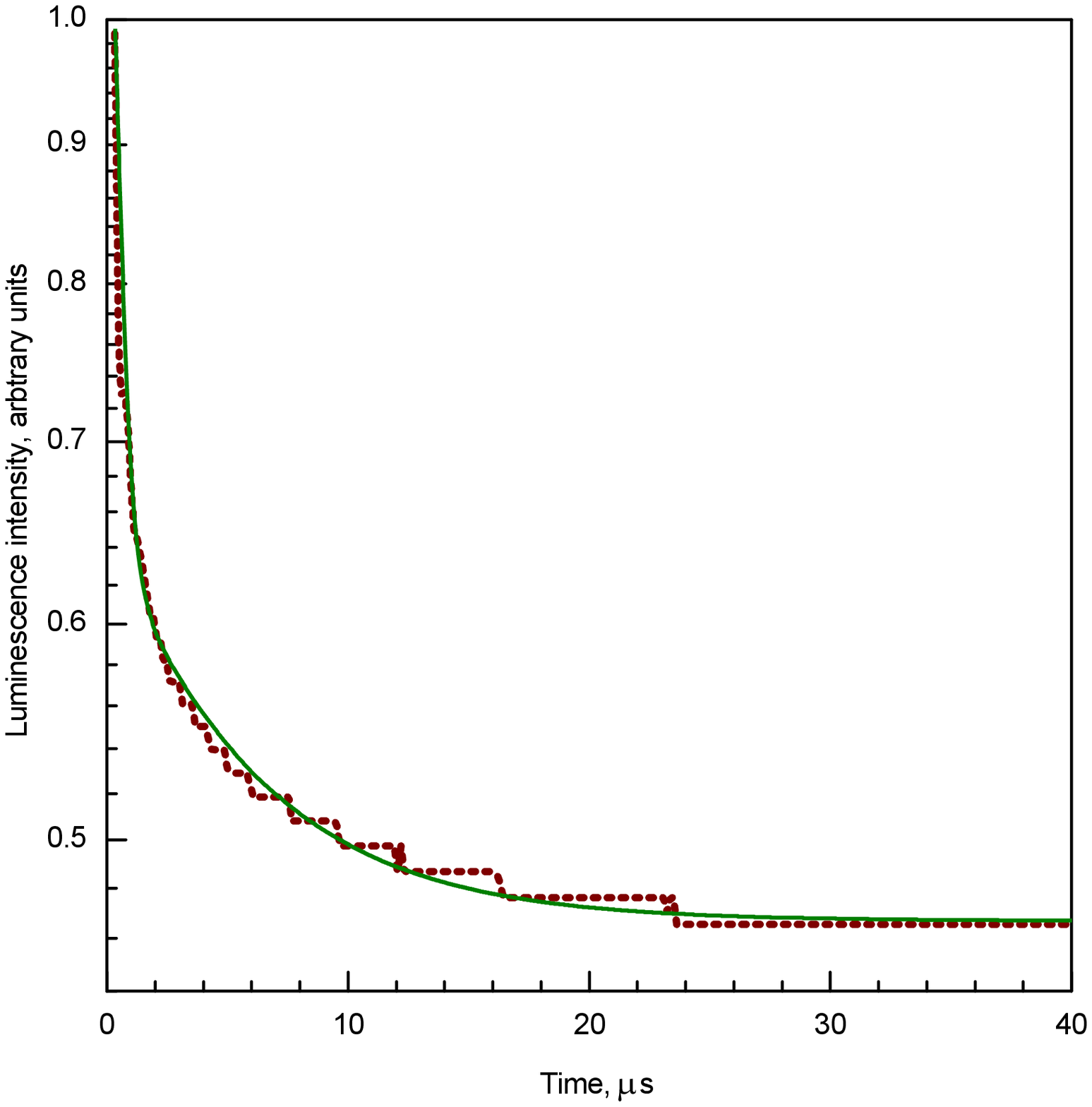}
\label{fig:3A}
}
\subfigure[]{%
\includegraphics[width=8.75cm, bb=70 290 550 770]{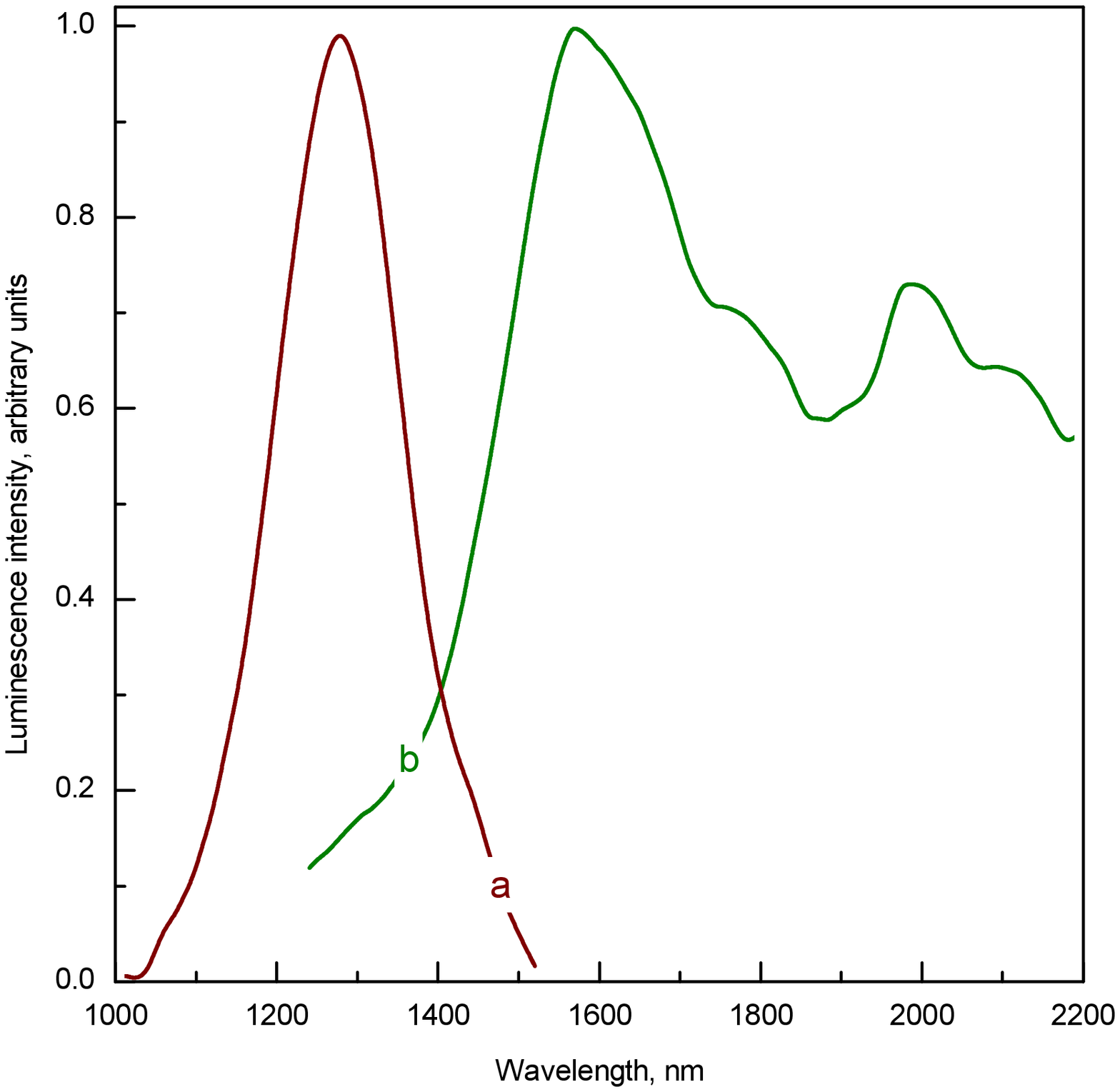}
\label{fig:3B}
}
\caption{%
\subref{fig:3A}.~Luminescence decay curve (As$_{10}$Ge$_{35}$S$_{55} +
0.45$\%~Bi, excitation at 975~nm): $\tau_1 \approx 0.3$~\mks, $\tau_2 \approx
5.7$~\mks. \quad
\subref{fig:3B}.~Luminescence spectra of
(a)~80GeS$_{2}$--20Ga$_2$S$_{3} + 0.5$~at.\%~Bi glass (excitation at 839~nm);
(b)~As$_{10}$Ge$_{35}$S$_{55} + 0.5$~at.\%~Bi glass (excitation at 975~nm, the
spectra are normalized to 1).
}
\label{fig:3}
\end{figure*}

The transmission spectra of glass samples with different composition are shown
in Fig.~\ref{fig:1}. One can see how the short-wave transmission edge depends on
the glass composition and Bi content. Increasing the Bi concentration shifts the
short-wave edge to the IR region (Fig.~\ref{fig:1}(\subref{fig:1A})). The
GeS$_{1.5}$:Bi sample has a lower transmission than the GeS$_{1.35}$ sample
without Bi (Fig.~\ref{fig:1}(\subref{fig:1B})). This fact may be due to the
refractive index  increase and a formation of inhomogeneities due to a partial
crystallization. Addition of As expands the transmission range and increases the
transmission (Fig.~\ref{fig:1}(\subref{fig:1C})), which may be related to an
increase in resistance to crystallization. In the transmission spectra of the
samples with Bi there are no distinctly expressed absorption bands.

\mbox{Ga--Ge--S} system glasses doped with Bi were also studied. Bulk samples of
such glasses easily crystallize even when they are cooled in cold water.
Therefore it is extremely difficult to make optical fibers using this glass
composition. \mbox{Ga--Ge--S} system glasses are more transparent in the visible
range than \mbox{As--Ge--S} and \mbox{Ge--S} system glasses, but Bi addition
makes them far more less transparent. The absorption edge is significantly
shifted towards shorter wavelengths in such doped glasses
(Fig.~\ref{fig:1}(\subref{fig:1D})).

The luminescence spectra of glass samples with different composition are shown
in Fig.~\ref{fig:2}. \mbox{Ge--S} glasses demonstrate a relatively low
luminescence band intensity, but its width covers a wide spectral range
(Fig.~\ref{fig:2}(\subref{fig:2A})). GeS$_{1.5}$ glass has a higher luminescence
intensity than GeS$_{1.35}$. We couldn't obtain Bi-doped \mbox{Ge--S} glass with
higher S content because of too strong a tendency to crystallization. It is
known that the well-studied \mbox{As--S} glass system shows no tendency to
crystallization, and it was decided to add As into the \mbox{Ge--S} system to
increase the glass-forming ability of \mbox{Ge--S} system glasses. The
\mbox{S/Ge} ratio was no higher than 2. Also As addition resulted in a
significant increase in the luminescence intensity
(Figs.~\ref{fig:2}(\subref{fig:2A}) and \ref{fig:2}(\subref{fig:2B})).

Measurements of the concentration series of samples
(Figs.~\ref{fig:2}(\subref{fig:2B}) and \ref{fig:2}(\subref{fig:2C})) with
various proportion of As and Bi have shown that the maximum intensity of the IR
luminescence band occurs in the As$_{10}$Ge$_{35}$S$_{55} + 0.5$~at.\%~Bi glass.
Slightly lower intensity is observed in As$_{20}$Ge$_{30}$S$_{50} +
0.5$~at.\%~Bi glass.

Arsenic presumably affects mainly the resistance to crystallization and the
intensity of luminescence, but does not change its spectral structure. Arsenic
unlikely leads to the formation of new centers of luminescence absent in
\mbox{Ge--S}.

Similar luminescence bands were also detected in optical fibers made of the same
glass (Fig.~\ref{fig:2}(\subref{fig:2D}), to be compared with
Fig.~\ref{fig:2}(\subref{fig:2C})).

The luminescence lifetime was about 6~\mks{}
(Fig.~\ref{fig:3}(\subref{fig:3A})).

In \mbox{Ga--Ge--S} glasses the IR luminescence band maximum is located near
1.28~\mkm{} (Fig.~\ref{fig:3}(\subref{fig:3B})), and the band shape is close to
Gaussian with FWHM about 200~nm. Band intensity is comparable to that in
\mbox{As--Ge--S} glasses. Disadvantages of these glass systems compared with
\mbox{As--Ge--S} glasses consist in much higher tendency to crystallization and
a relatively narrow IR luminescence band.

\section{Modeling of Bi-related luminescence centers}
To understand the origin of Bi-related centers responsible for the IR
luminescence observed in GeS$_{2-x}$ glasses and based on assumptions on the
role of subvalent bismuth states \cite{Peng11}, we performed computer simulation
of the structure and absorption spectra of several centers most probably formed
by Bi atoms in GeS$_{2-x}$ glass network. Namely, trivalent (\Bipiii) and
divalent (\Bipii) substitutional Bi centers, interstitial \Bipi{} ions,
interstitial \Biz{} atoms, and complexes formed by an interstitial Bi atom and
an S vacancy were studied. The structure of the centers was calculated using
network models with periodic boundary conditions. The unit cell of defect-free
GeS$_2$ network containing 24 GeS$_2$ groups was prepared by ab initio
(Car-Parrinello) molecular dynamics. To study Bi-related centers, a Bi atom was
placed in the central part of the unit cell. S vacancies were formed by removal
of one of the S atoms. The charge state of a center was determined by the total
unit cell charge. Equilibrium configurations of Bi-related centers were found by
a complete geometry optimization with the gradient method in the plane wave
basis using the generalized gradient approximation of the density functional
theory and pseudopotentials. All the structure-related calculations were
performed with the \QE{} package \cite{QE}. The configurations of Bi-related
centers were used further to calculate the absorption spectra of the centers
using the Bethe–Salpeter equation method  based on the all-electron
full-potential linearized augmented plane wave approach taking into account the
spin-orbit interaction essential for Bi-containing systems. The \Elk{} code
\cite{Elk} was used in spectra calculations. To estimate the luminescence Stokes
shift, the configurational coordinate diagrams of Bi-related centers were
studied in the frame of a simple model. The modeling is described in detail in
Refs.~\cite{Sokolov13a, Sokolov13b}.
\begin{figure}
\begin{center}
\includegraphics[width=8.55cm, bb=105 185 503 615]{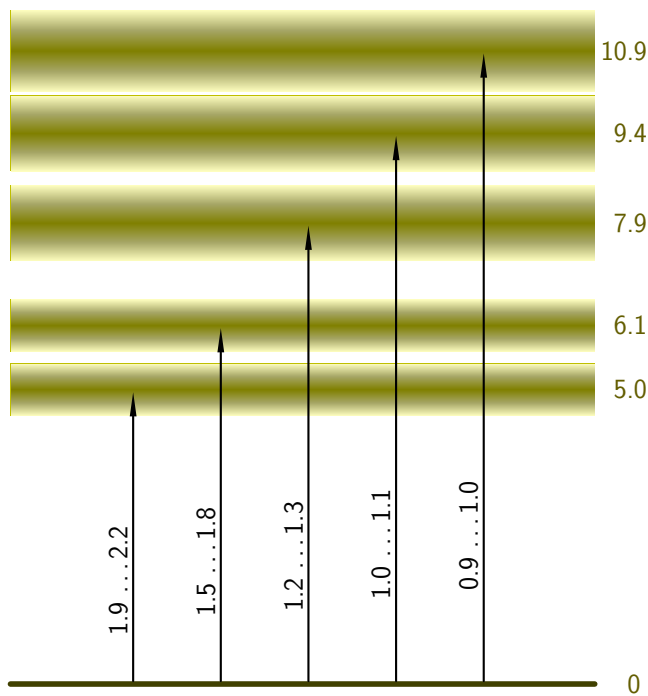}
\end{center}
\caption{%
Calculated levels and transitions in the complex formed by an interstitial Bi
atom and an S vacancy (energy in $10^3$~\cminv, transition wavelengths in
\mkm).
}
\label{fig:4}
\end{figure}

The modeling shows that threefold and twofold coordinated Bi atoms bonded with
Ge atoms by three or two bridging S atoms, respectively (\Bipiii{} or \Bipii{}
centers), are very likely to occur in GeS$_2$ network. In such centers bismuth
is in three- and divalent state, respectively. Formation energy estimations
based on the calculated total energies of corresponding unit cells show that
\Bipiii{} centers are favorable. Calculated wavelengths of the absorption bands
of such \Bipiii{} and \Bipii{} centers in GeS$_2$ network agree well with
the experimental data available on the spectral properties of similar Bi-related
centers in several hosts \cite{Blasse68, Blasse84, Blasse94, Blasse97,
Srivastava98}. These centers are not related to IR luminescence and are of no
interest for our investigation, since both the absorption and luminescence bands
of the centers fall in the host self-absorption range.

According to the calculations, both \Biz{} atoms and \Bipi{} ions are not stable
in interstitial positions of GeS$_2$ network. In the regular network
interstitial Bi atoms tend to form bonds with the surrounding S atoms giving
rise to network rearrangement with the above-mentioned \Bipiii{} and \Bipii{}
substitutional centers formed. The energy gain of such a rearrangement is
estimated to be about 0.9~eV per unit cell. Therefore in GeS$_{2-x}$ based
glasses the IR luminescence is unlikely to be caused by interstitial \Bipi{}
ions. Notice that the characteristic luminescence band of \Bipi{} ions is known
from the experimental and theoretical data to be located near 1~\mkm{}
\cite{Sokolov13a, Sokolov13b, Davis67}.

\setatomsep{2.3em}
On the other hand, the calculations show that if an interstitial Bi atom occurs
in GeS$_2$ network in vicinity of such an intrinsic defect as a S vacancy,
\GeGe, a complex of the interstitial \Biz{} atom with the vacancy,
\schemestart[][south west]
\chemfig{{}~[,0.56]Ge?-[:25]Bi-[:-25]Ge?~[,0.56]{}}
\schemestop,
is formed instead of the \Bipiii{} and \Bipii{} substitutional centers. The
total energy of the unit cell with such a complex is estimated to be about
0.6~eV lower than that of the unit cell with separated \Bipiii{} (or \Bipii)
center and S vacancy. In this complex center bismuth turns out to be in the
state close to monovalent state due to redistribution of electron density.
Effective electronic charge $\approx - 0.8\left|\textrm{e}\right|$ is displaced
from Bi and Ge atoms towards the area between Bi and Ge atoms, and to a lesser
extent into the area between both Ge atoms. Thus a three-center system is formed
consisting of a Bi atom and two Ge atoms with a coordination-type bonding
between three atoms. In a rough approximation, basing on the described electron
density redistribution, the complex may be considered as a pair of charged
centers, the interstitial positively charged ion, \Bipi, adjacent to the
negatively charged S vacancy, \mbox{$\equiv\!\textrm{Ge}
{\overset{{}^\bullet}{\relbar}} \textrm{Ge}\!\equiv$}. This make it possible to
describe the electronic structure of the center in terms of a crystal field
model similar to ones used previously for Bi-related centers in TlCl:Bi, CsI:Bi
\cite{Sokolov13a}, SiO$_2$:Bi and GeO$_2$:Bi \cite{Sokolov13b}, and in AgCl:Bi
\cite{Plotnichenko13}. The ground state and the first two excited states of the
\Bipi{} ion are known to arise from the triplet state, \Term{3}{P}{}{},
(electron configuration 6s$^2$p$^2$) split by strong intra-atomic spin-orbit
interaction in three components, the ground state, \Term{3}{P}{0}{}, and excited
states, \Term{3}{P}{1}{}{} and \Term{3}{P}{2}{}, with the energies of about
13300 and 17000~\cminv{} in a free ion, respectively. These states can be
further split and mutually mixed under the influence of the crystal field of the
\Bipi{} ion environment in the glass network. For the "Bi --- vacancy" complex
such an axial crystal field caused by a charged vacancy turns out to be strong.
In an axial crystal field the ground state of \Bipi{} is not split, and the
\Term{3}{P}{1}{}{} and \Term{3}{P}{2}{}{} excited states are split into two and
three levels, respectively. The dipole transitions between the ground and
excited states forbidden in the free ion become weakly allowed due to mixing of
the wave functions. A relatively small (in comparison with Bi-related centers in
other hosts) IR luminescence lifetime can be explained by a relatively high
degree of \Bipi{} ion perturbation by the vacancy crystal field and,
respectively, a significant state mixing. For example, our calculations
\cite{Sokolov13b} similar to the present ones have shown that in the complex
center formed by interstitial Bi atom and anion vacancy in the GeS$_2$ network
the bonding is noticeably stronger than in similar centers in SiO$_2$ or GeO$_2$
networks.

The calculated levels and transitions in the complex center formed by the
interstitial Bi atom and the S vacancy are shown in Fig.~\ref{fig:4}. The Stokes
shift of the luminescence is found to be small, at least for the longest-wave
transitions. This can be explained by relatively small admixture of Ge
electronic states to the wave functions of this complex, so that small
displacement of the Bi atom does not result in a noticeable change of electronic
states of the complex. Thus, one expects the "Bi --- vacancy" complex in the
GeS$_2$ network to cause IR luminescence bands near $1.9 \ldots 2.1$~\mkm{} and
$1.5 \ldots 1.8$~\mkm, when excited both at the same absorption wavelength and
in three absorption bands in the $0.9 \ldots 1.3$~\mkm{} range.

\section{Conclusion}
Basing on the above-mentioned total energy estimations it may be concluded that
bismuth occurs in regular GeS$_2$ network mainly as trivalent (\Bipiii)
substitutional centers, and, probably, as divalent (\Bipii) ones. However,
centers formed by interstitial Bi atoms and S vacancies would be expected to
occur in sulfur-deficient network, and only such complexes may give rise to the
bismuth-related IR luminescence. Such a center may be considered as an
interstitial \Bipi{} ion adjacent to negatively charged S vacancy. Comparison of
the calculation results with the experimental data allow us to suggest that
these complexes make the main contribution to the IR luminescence in
GeS$_{2-x}$:Bi glasses.

Arsenic addition to \mbox{Ge--S} glass significantly enhances its resistance to
crystallization and made it possible to draw optical fibers. Arsenic also
significantly increased the intensity of luminescence. Such an increase in
luminescence intensity in the arsenic-containing glasses can be explained by a
decrease in concentration of Bi ions with oxidation degree higher than 1 due to
reduction properties of arsenic. Furthermore, arsenic occurs in the glass
network mainly in the form of threefold atoms and prevents the formation of the
above-described substitution centers \Bipiii{} and \Bipii{} centers. Sulfur
deficiency in \mbox{As--Ge--S} compositions may promote the formation of S
vacancies and complexes with interstitial Bi atoms.

\section*{Acknowledgments}
The authors are grateful to Dr.~B.~I.~Galagan for his help in luminescence
lifetime measurements and for valuable discussions. This work is supported in
part by Basic Research Program of the Presidium of the Russian Academy of
Sciences and by Russian Foundation for Basic Research (grant
\mbox{12-02-00907}).

%
% References %%%%%%%%%%%%%%%%%%%%%%%%%%%%%%%%%%%%%%%%%%%%%%%%%%%%%%%%%%%%%%%%%%%
%

\end{document}